\documentclass[%
 reprint,
 amsmath,amssymb,
 aps,
pra,
]{revtex4-1}

\usepackage{graphicx}
\usepackage{multirow}
\usepackage{dcolumn}
\usepackage{bm}
\usepackage{color}

\newcommand{\hide}[1]{}

\newcommand{\fig}[1]{Fig.\,\ref{#1}}

\newcommand{\nofig}[1]{(\ref{#1})}

\begin{document}

\preprint{APS/123-QED}

\title{Optimal population transfer in combined Feshbach Resonance and STIRAP process}

\author{Phillip Price}
\affiliation{%
 Physics Department, University of Connecticut, Storrs, CT 06269, USA
}%

\author{S. F. Yelin}%
\affiliation{%
 Physics Department, University of Connecticut, Storrs, CT 06269, USA
}%
\affiliation{%
 Physics Department, Harvard University, Cambridge, MA 02138, USA
}%

\date{\today}

\begin{abstract}
We present a method for the creation and control of cold molecules that involves coherently combining Feshbach Resonances and STIRAP.  We present analytical and numerical results showing how to optimize this process that can be implemented using techniques readily available in standard experimental setups.  This will provide a link in the chain from atoms to ground state molecules and can serve as a building block towards more complex processes in coherent ultracold chemistry.
\end{abstract}

\maketitle


\section{\label{sec:Intro}Introduction}
Ultracold molecules have many potential uses in prominent areas such as quantum computation \cite{earlySTIRAP2}, control of chemical reactions \cite{Danzl1062, earlySTIRAP, RAPrev, RAPrev2}, fundamental measurements \cite{DblSTIRAP}, and few-body collision physics \cite{Danzl1062, earlySTIRAP2}.  Their rich internal energy structure that makes them useful in these applications is the same property that makes creating these objects difficult.  
One standard method for creating these ultracold molecules involves first creating ultracold atoms, then using incoherent procedures such as magneto-association, also known as Feshbach Resonance (FR) \cite{earlyFR, earlyFR2, FRrev1, FRrev2, FeshRes}, to form vibrationally high energy molecules, and afterwards transfer the molecules into their ground state via STIRAP \cite{earlySTIRAP, earlySTIRAP2, earlySTIRAP3, generalSTIRAP}. One of the side effects of this methodology is that the intermediate state, the vibrationally hot molecules, sit and wait for the incoherent process to end before being cooled using coherent techniques.  This opens the door to lose population through environmental factors, such as collisions or decay to outside states.  Taking inspiration from double STIRAP procedures \cite{DblSTIRAP, Danzl1062} we look to coherently combine the process of magneto-association and STIRAP in an attempt to minimize the time these molecules remain in this unstable state.  By chaining these two processes together in a coherent manner we create a toolkit for true deterministic and coherent ultracold chemistry.
In this paper we describe how we approached this task and the numerical optimization that goes into creating an ideal case.  The major shift in framework we made is viewing the process of magneto-association, otherwise known as a Feshbach Resonance (FR), through the lens of the well understood coherent process of Rapid Adiabatic Passage (RAP) \cite{RAPrev, RAPrev2, earlyRAP, earlyRAP2, earlyRAP3}.  With that shift, we attempt to find a suitable dark state for this process and optimize accordingly.  Ideally this technique can be used to chain together other elementary processes in this Lego-type coherent fashion.

\section{\label{sec:MDesc}Model Description}

Our model describes a continuous, coherent process from ultraccold, separated atoms into ultracold molecules.  Starting with separate atoms, we sweep through a Feshbach resonance to create a Feshbach molecule, i.e. a molecule in a highly excited vibrational state of the electronic ground state.  Then we apply STIRAP to the resulting molecule to transfer the molecule into a lower energy state.  The key factor that makes this method novel and useful compared to standard techniques is the coherent nature of the process.  While common experimental techniques involve storing the Feshbach molecules in an optical trap and then performing STIRAP to cool them, our method is done in one continuous sweep to reduce the time the molecule spends in unstable intermediary states.  Figure 1 shows a visualization of this process.

\begin{figure}[htbp]
\begin{center}
\includegraphics[width=0.95\linewidth]{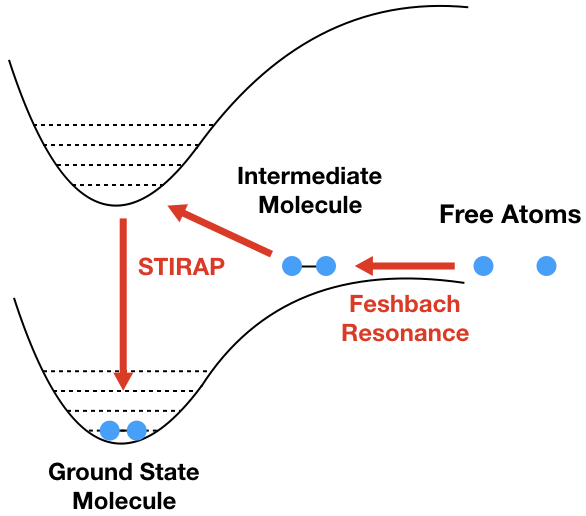}
\caption{General overview of the process, working from right to left.  A collection of ultracold atoms goes through a Feshbach resonance to create ultracold molecules.  Then, without storing these high energy molecules, we coherently apply STIRAP to take the molecule into a more stable ground state level. Picture inspired by \cite{pictureSource}. }
\label{fig:mpc}
\end{center}
\end{figure}

\begin{figure}[htbp]
\begin{center}
\includegraphics[width=0.95\linewidth]{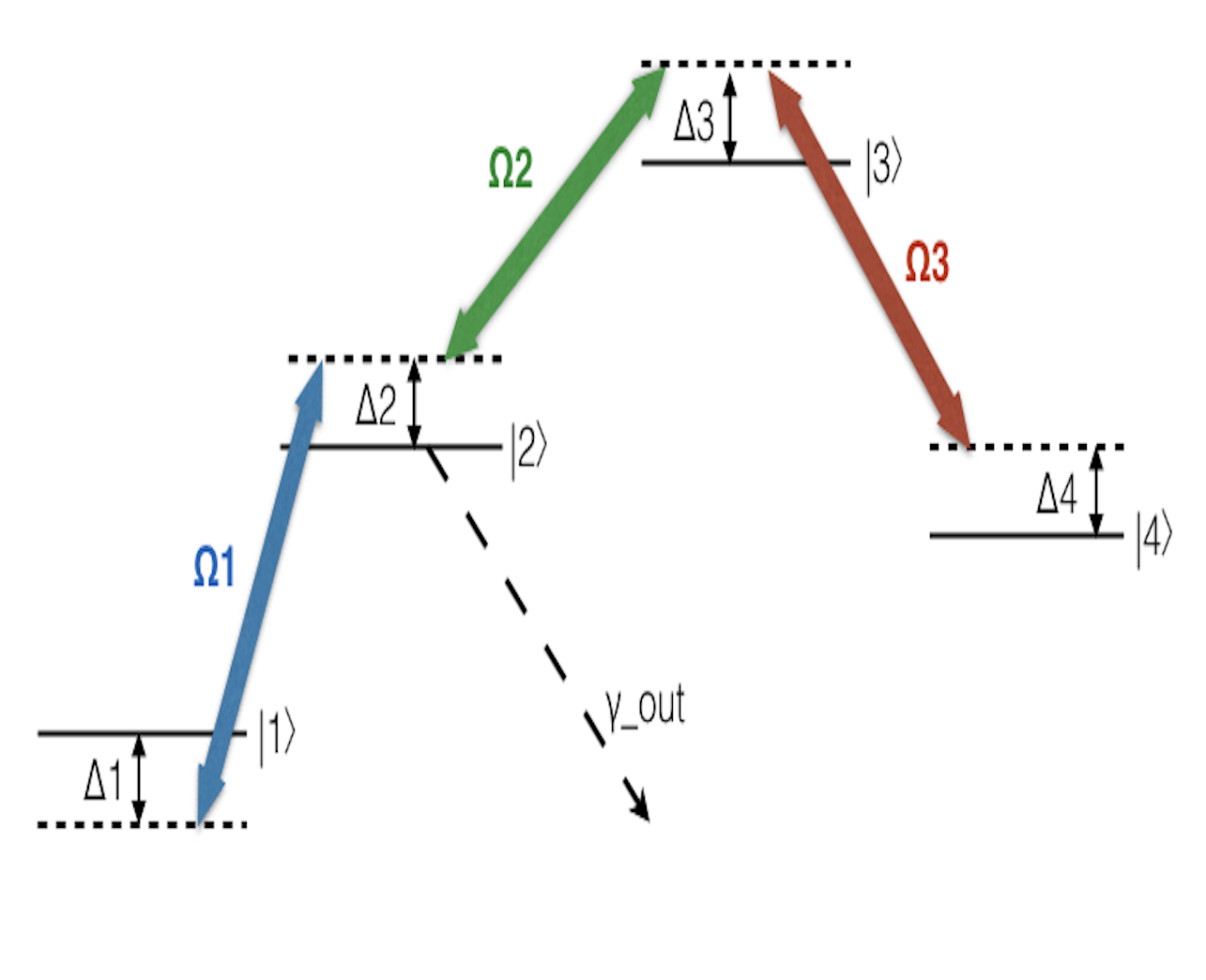}
\caption{ Energy level scheme for the 4-level model of the system.  $ \Omega 1 $ is related to the width of the Feshbach resonance, while $\Omega 2$ and $\Omega 3$ are the pump and Stokes pulses respectively in a standard 3 level STIRAP procedure.  There are 3 unlabeled decay rates, from $\vert 2 \rangle$ into $\vert 1 \rangle$, from $\vert 3 \rangle$ into $\vert 2 \rangle$, and from $\vert 3 \rangle$ into $\vert 4 \rangle$.}
\label{fig:4lvls}
\end{center}
\end{figure}

\subsection{\label{sec:MCon}Model construction}
Figure 2 is a visualization of the level scheme for this combined FR into STIRAP process.  $\vert 1 \rangle$ represents the state of separated ultracold atoms.  $\vert 2 \rangle$ represents the state of the newly combined Feshbach molecules.  $\vert 4 \rangle$ represents our end state of ground state molecules, and $\vert 3 \rangle$ represents the intermediate molecular state of a standard STIRAP procedure.  From this we get the following Hamiltonian for our four level system : 

\begin{equation}
H = \left[ \begin{array}{c c c c} \Delta_1 & \Omega_1 & 0 & 0 \\
\Omega_1 & \Delta_2 & \Omega_2 & 0 \\
0 & \Omega_2 & \Delta_3 & \Omega_3 \\
0 & 0 & \Omega_3 & \Delta_4
\end{array}
\right]
\end{equation}

The $\Delta$'s are detunings of levels from their bare atomic resonances.  Our investigation will differ from other investigations into four-level systems \cite{4lvlSTIRAP, 4lvlSTIRAPcomputations} in that we do not apriori assume certain detunings are identically zero or automatically enforce any n-photon resonances.  In addition to the decay from $\vert 2 \rangle$ out of the system, we have 3 other decay parameters.  We have $\gamma_{s1}$ and $\gamma_{s2}$ from $\vert 3 \rangle$ into $\vert 2 \rangle$ and $\vert 4 \rangle$ respectively, as well as $\gamma_{f}$ from $\vert 2 \rangle$ into $\vert 1 \rangle$.   
This model does not take into account any of the inefficiencies associated with standard FR processes, so population numbers reflect only the percentage of Feshbach molecules that are actually created via the initial magnetic field sweep, which typically is about 20\% of the ultracold atom population.   
We also neglect to include decay from $\vert 3 \rangle$ out of the system.  From our numerical solutions the population spends negligible time in $\vert 3 \rangle$ and thus extra decay from this level would have no effect.  

\subsection{\label{sec:FR}Feshbach resonance as rapid adiabatic passage}

To justify this model we treat FR as RAP.  The Hamiltonian for a coupled two-channel model of FR is \cite{FRrev1}

$$
H_{tc} = \left[ \begin{array}{c c} H_{bg} & W(r) \\
W(r) & H_{cl}(B)
\end{array}
\right]
$$

The standard Hamiltonian for RAP is 

$$
H_{RAP} = \left[ \begin{array}{c c} 0 & \Omega \\
\Omega & \Delta
\end{array}
\right]
$$

According to \cite{FRrev1} we can reduce the diagonal elements of $H_{tc}$ to just the difference between the entrance and closed channels of the system, i.e. $H_{bg} \rightarrow 0$ and $H_{cl} \rightarrow \left( H_{cl}(B) - H_{bg} \right)$.  The optical detuning ($\Delta$) of an RAP process is analogous to the magnetic detuning ($B$) in a FR process, while the Rabi frequencies ($\Omega$) in RAP serve the same purpose as the level couplings ($W(r)$) of FR in determining the width of the resonance. The similarity of both processes can be seen in their energy dynamics.  This is shown in the appendix.

\section{\label{sec:QA}Qualitative Analysis}
Our goal is to maximize the coherent transfer of population from $\vert 1 \rangle$ into $\vert 4 \rangle$.  An easy way to do this would be to find a dark state of the Hamiltonian that has components of both $\vert 1 \rangle$ and $\vert 4 \rangle$.  Then we could adiabatically manipulate the parameters of our system such that the state initially is aligned along $\vert 1 \rangle$ and ends fully aligned along $\vert 4 \rangle$.  This is the bare minimum we are looking for at first.  Once those conditions are satisfied we would like to impose further conditions of minimizing the amount of population that is in $\vert 2 \rangle$ and $\vert 3 \rangle$ throughout the process, as those states are assumed to be the most unstable and prone to decays out of the system.  The first step in finding a dark state is looking at the determinant of the Hamiltonian

\begin{multline} \label{eq:detH}
\det H  = \Delta_1 \Delta_2 \Delta_3 \Delta_4 - \Delta_3 \Delta_4 {\Omega_1}^2 - \Delta_1 \Delta_4 {\Omega_2}^2 \\
  - \Delta_1 \Delta_2 {\Omega_3}^2 + {\Omega_1}^2 {\Omega_3}^2
\end{multline}

This is not identically zero, so no dark state exists apriori in this system.  To move forward we artificially create one by enforcing certain relations between the parameters.  First, note that this expression can be simplified by taking advantage of the freedom to set the zero energy of this system anywhere.  Setting one of the detunings to 0 will negate some terms in this expression and make it more manageable.  We choose to set $\Delta_4 = 0$, which reduces the complexity of the determinant to the last two terms in \eqref{eq:detH}.  These two terms cancel under following relation

\begin{equation} \label{eq:DSC}
{\Omega_1}^2 = \Delta_1 \Delta_2 
\end{equation}

With \eqref{eq:DSC} the determinant of our Hamiltonian becomes 0 and the following unnormalized dark state appears

\begin{equation} \label{eq:DarkState}
\vert \Psi \rangle = \Omega_1 \Omega_3 \vert 1 \rangle - \Delta_1 \Omega_3 \vert 2 \rangle + \Delta_1 \Omega_2 \vert 4 \rangle 
\end{equation}

As desired this dark state has components along $\vert 1 \rangle$ and $\vert 4 \rangle$.  While there is an additional component along $\vert 2 \rangle$, there is no component along $\vert 3 \rangle$.  Getting this dark state to initially line up with $\vert 1 \rangle$ is difficult primarily because of the dark state condition \eqref{eq:DSC}.  Our approximation is to start with $\Omega_1$ and $\Omega_2$ equal to $0$, with $\Omega_3$ and $\Delta_2$ both nonzero.  $\Delta_3$ plays no significant role in this discussion, and $\Delta_4$ is already $0$ by assumption.  Our argument for this setup is that while technically all of the dark state components would be zero with these assignments, the components along $\vert 2 \rangle$ and $\vert 4 \rangle$ will be more strongly $0$ than the component along $\vert 1 \rangle$.  The component along $\vert 1 \rangle$ is a nonzero number multiplied by something very close to zero, whereas the components along $\vert 2 \rangle$ and $\vert 4 \rangle$ are the multiplication of two numbers very close to zero.  At the end of our process we want the dark state to lie along $\vert 4 \rangle$, so in a similar fashion we choose $\Omega_2$ and $\Omega_1$ to be nonzero while $\Omega_3$ and $\Delta_2$ become $0$ or close to it.  
Other choices can be made for the detunings to set to 0, but they do little to simplify the problem.  Setting $\Delta_1 = 0$ results in a dark state with components along $\vert 1 \rangle$, $\vert 3 \rangle$, and $\vert 4 \rangle$ while setting the other two detunings to zero results in a dark state with components along all four bare states.
Throughout this procedure it will be useful to look at the coupling strengths between our dark state and other states in which we wish to avoid large populations.  First we move into an adiabatic basis.  We choose our dark state \eqref{eq:DarkState} and the excited state $\vert 3 \rangle$ as two of the new basis vectors. then use Gram-Schmidt orthogonalization to get the remaining two basis states, which we will call bright states.  The transformation matrix $R$ then becomes 

$$
R = \left[
\begin{array}{cccc}
\Delta_1 & - \Delta_1 \Omega_1 \Omega_2 & 0 & \Omega_1 \Omega_3 \\
\Omega_1 & {\Delta_1}^2 \Omega_2 & 0 & -\Delta_1 \Omega_3 \\
0 & 0 & 1 & 0 \\
0 & \left( {\Delta_1}^2 + {\Omega_1}^2 \right) \Omega_3 & 0 & \Delta_1 \Omega_2
\end{array}
\right]
$$

where columns 1 and 2 are the bright states, column 3 is the excited state, and column 4 is our dark state.  Sine the parameters in this matrix are time-dependent, this time-dependent transformation leads to the following transformed Hamiltonian
\begin{equation} \label{eq:Hprime}
H^{\prime} = R H R^{\dag} - \imath R^{\dag} \dot{R}
\end{equation}
The coupling between the dark state and the excited state is identically 0 in this transformation.  Therefore in our adiabatic considerations we focus on the coupling between the dark state and the two constructed bright states.  By minimizing this coupling we should be able to stay in our dark state even in the presence of incoherent processes like decay in the system \cite{Wu16}

\section{\label{sec:NumRes}Numerical Results}

In order to optimize the process and its timing, we employ numerical calculations. While the results shown here are not systematically optimized, combinations of intensity, detuning, time delays and processing speeds are shown here to demonstrate the power of the procedure compared to typical experimental setups. 

A typical functional form of the Rabi frequencies that optimally maintain the coherence necessary for STIRAP consists of  {\it arctan}-functions:
 $$
  \Omega_i (t) = \frac{\Omega_{mag}}{\pi} \left( \frac{\pi}{2} \pm \arctan \left( \frac{t - t_0}{\tau}  \right) \right),
 $$
where $\Omega_{mag}$, $t_0$, and $\tau$ are the numerical parameters that form the search space.  The form is such that for $- (+)$ the $\Omega$'s start (end) at 0 and end (start) at $\Omega_{mag}$.  We also take a similar form for $\Delta_2$, where $\Delta_2$ starts near its maximal value and ends near $0$.  $\Delta_1$ is numerically determined by the dark state condition \eqref{eq:DSC}.

\begin{figure}[htbp]
\begin{center}
\includegraphics[width=0.95\linewidth]{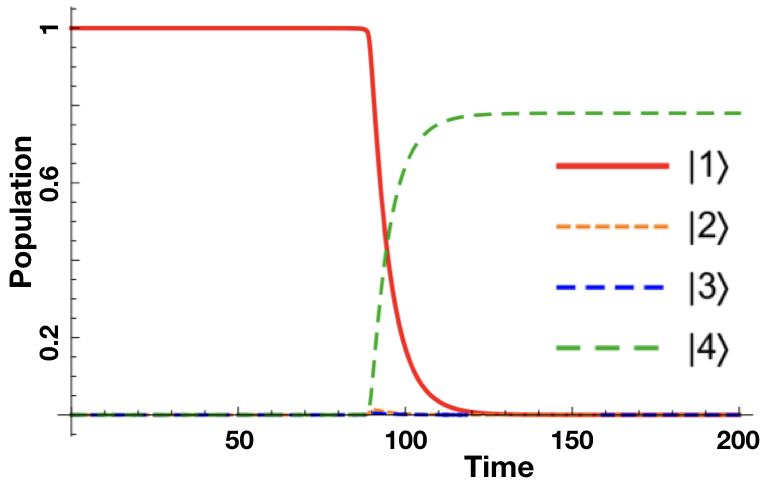}
\caption{A typical time evolution of the population in each of the four bare states throughout the process.   }
\label{fig:pops}
\end{center}
\end{figure}

Setting the decay rates $\gamma_f = \gamma_{out} = 2$ and $\gamma_{s1} = \gamma{s2} = 20$ we found a typical evolution , represented by \fig{fig:pops}.  

\begin{figure}[htbp]
\begin{center}
\includegraphics[width=0.95\linewidth]{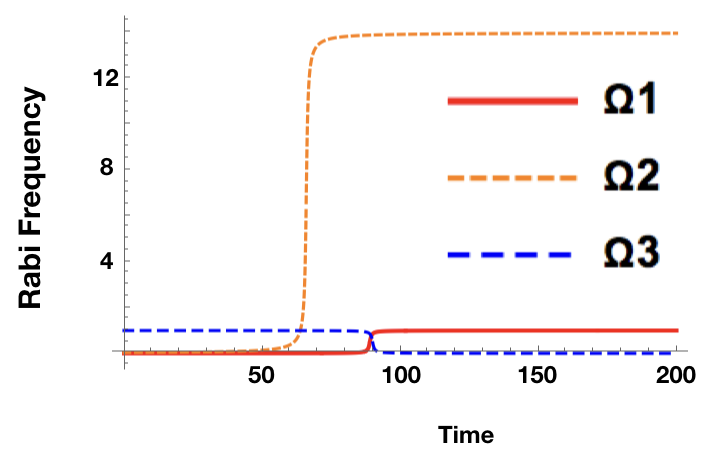}
\caption{Rabi frequencies ($\Omega$'s) for time evolution in \fig{fig:pops}.}
\label{fig:omegas}
\end{center}
\end{figure}

\begin{figure}[htbp]
\begin{center}
\includegraphics[width=0.95\linewidth]{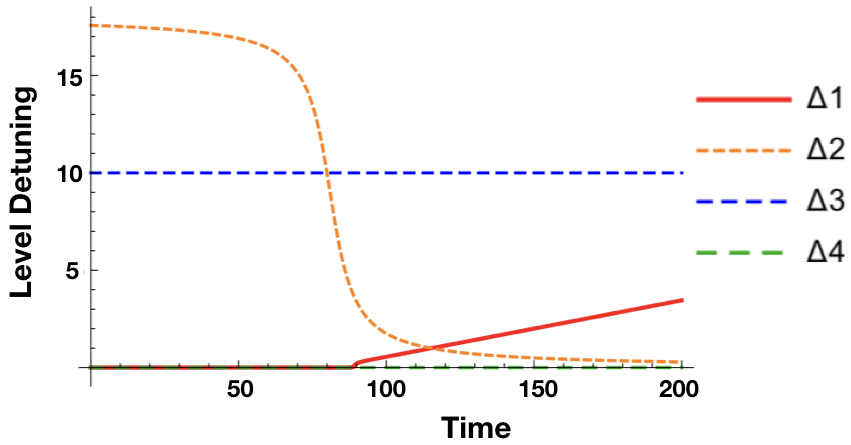}
\caption{Relevant level detunings for time evolution presented in \fig{fig:pops}.}
\label{fig:deltas}
\end{center}
\end{figure}

Figures \nofig{fig:omegas} and \nofig{fig:deltas} show the Rabi frequencies and detunings respectively needed to achieve this result.  Note that the decay rates are near the magnitude of the Rabi frequencies needed to achieve this result.  To help understand what these numbers mean, it's useful to compare them with physically realizable parameters.  Scaling the decay rate $\gamma_{s1}$ to the physically relevant quantity $12.5 \times 10^9 s^{-1}$ results in a maximal Rabi frequency of $8.75 \times 10^9 s^{-1}$ for $\Omega_{2}$ \cite{4lvlSTIRAPcomputations}.

\begin{figure}[htbp]
\begin{center}
\includegraphics[width=0.95\linewidth]{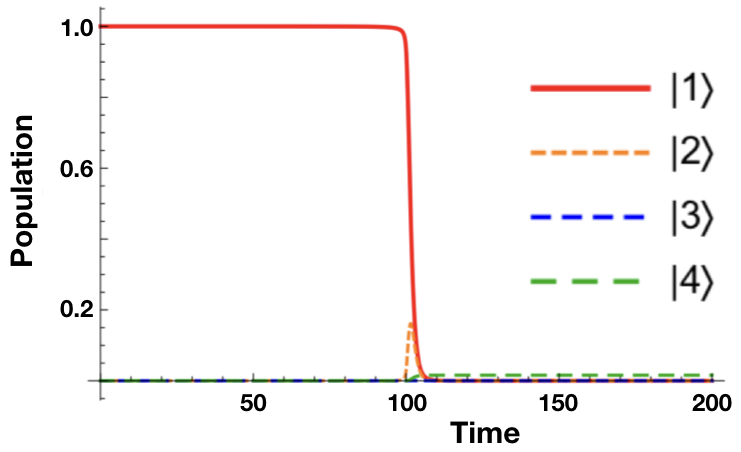}
\caption{Time evolution of populations with FR and STIRAP as separate processes when the decay of state $\vert 2 \rangle$ is not vanishingly small.  This was optimized in a similar way as \fig{fig:pops}, using the same decay rates but varying the strength and time-scale of the Rabi frequencies and level detunings.}
\label{fig:separate}
\end{center}
\end{figure}

This is a considerably better result than what can be achieved by treating the same four level system as two separate coherent processes of RAP followed by STIRAP, as seen in \fig{fig:separate}.  This is a massive improvement for a medium-strength decay out of the system.  While very strong decay out of the system will, up to now, produce similar results, this new method shows much improvement over treating these two processes independently.  

Results for different combinations of system parameters (in particular, different strengths of decay out of state $\vert 2 \rangle$) are shown in .

\section{\label{sec:Conc}Conclusion}

This work provides a novel and easily testable framework for the creation of ultracold molecules.  In addition, this coherent combination of the two steps can, in principle, be inserted anywhere needed in a chain of steps, joining together any resonant and STIRAP processes.  Current experimental setups should be able to incorporate these ideas fairly easily, as the proposed method does not call for new equipment, merely an adjustment of standard techniques by adjusting to time dependent Rabi frequencies and detunings.  One issue this work does not address yet is the relative inefficiency of the resonant process, particularly of Feshbach resonances. Due to the coherent nature of the STIRAP procedure, any combination of Feshbach resonance and STIRAP can only be used once, since repeated applications would empty the final state at the same rate as it would be filled. Further study into multiple iterations of this pulse scheme could mitigate that particular issue, altering this method to facilitate multiple runs that will capture more ground state molecules than are destroyed. This will be subject of an upcoming publication.  Further improvement on the method presented here could be achieved by taking advantage of the strong coupling between the intermediate levels $\vert 2 \rangle$ and $\vert 3 \rangle$ \cite{methodImprovement}.  

\section{Acknowledgements}

We would like to thank Robin C\^ot\`e and Florentin Reiter for helpful discussions.  We would also like to acknowledge funding from the National Science Foundation.

\appendix{}
\section{\label{sec:appA}Explicit Hamiltonian}
The entries from the transformed matrix $H^{\prime}$ mentioned in \eqref{eq:Hprime}

\begin{widetext}
\begin{align*}
{H^{\prime}}_{11} & = \Delta_1 + \frac{{\Omega_1}^2}{\Delta_1} \\
{H^{\prime}}_{22} & = 0 \\
{H^{\prime}}_{33} & = \Delta_3 \\
{H^{\prime}}_{44} & = 0 \\
{H^{\prime}}_{12} & = \frac{ \imath \Delta_1 \Omega_2 \left( \Delta_1 \dot{\Omega}_1 - \dot{\Delta}_1 \Omega_1 \right)}{\sqrt{{\Delta_1}^2 + {\Omega_1}^2} \sqrt{\left( {\Delta_1}^2 + {\Omega_1}^2 \right) \left( {\Omega_1}^2 {\Omega_3}^2 + {\Delta_1}^2 \left( {\Omega_2}^2 + {\Omega_3}^2 \right) \right)}  } \\
{H^{\prime}}_{13} & = \frac{\Omega_1 \Omega_2}{\sqrt{ {\Omega_1}^2 + {\Omega_2}^2}} \\
{H^{\prime}}_{14} & = \frac{ \imath \Omega_3 \left( \Omega_1 \dot{\Delta}_1 - \Delta_1 \dot{\Omega}_1 \right) }{ \sqrt{{\Delta_1}^2 + {\Omega_1}^2} \sqrt{ {\Omega_1}^2 {\Omega_3}^2 + {\Delta_1}^2 \left( {\Omega_2}^2 + {\Omega_3}^2 \right) } } \\
{H^{\prime}}_{23} & = \frac{ \sqrt{ \left( {\Delta_1}^2 + {\Omega_2}^2 \right) \left( {\Omega_1}^2 {\Omega_3}^2 + {\Delta_1}^2 \left( {\Omega_2}^2 + {\Omega_3}^2 \right) \right) } }{ {\Delta_1}^2 + {\Omega_1}^2 } \\
{H^{\prime}}_{24} & = \frac{ \imath \left( \Delta_1 \Omega_1 \Omega_2 \Omega_3 \dot{\Omega}_1 + {\Delta_1}^3 \left( \Omega_2 \dot{\Omega}_3 - \dot{\Omega}_2 \Omega_3 \right) - {\Omega_1}^2 \left( \Delta_1 \Omega_3 \dot{\Omega}_2 + \Omega_2 \left( \Omega_3 \dot{\Delta}_1 - \dot{\Omega}_3 \Delta_1 \right) \right) \right) }{ \sqrt{ {\Omega_1}^2 {\Omega_3}^2 + {\Delta_1}^2 \left( {\Omega_2}^2 + {\Omega_3}^2 \right)} \sqrt{ \left( {\Delta_1}^2 + {\Omega_1}^2 \right) \left( {\Omega_1}^2 {\Omega_3}^2 + {\Delta_1}^2 \left( {\Omega_2}^2 + {\Omega_3}^2 \right) \right) } } \\
{H^{\prime}}_{34} & = 0
\end{align*}
\end{widetext}

\section{Explicit parameter values}

\begin{tabular}{ c | c | c | c | c}
\hline
\multicolumn{2}{ c | } {Parameters} & \fig{fig:pops} & \fig{fig:Decay_Weak} & \fig{fig:Decay_Strong} \\ \hline
\multirow{3}{ 3em }{ $\Omega1$ } & $\Omega1_{mag}$ & 1 & 1 & 1 \\ 
 & $\Omega1_{offset}$ & 89 & 89 & 89 \\
 & $\Omega1_{tau}$ & 0.5 & 0.5 & 0.5 \\ \hline
\multirow{3}{ 3em }{ $\Omega2$ } & $\Omega2_{mag}$ & 14 & 14 & 14 \\ 
 & $\Omega2_{offset}$ & 66 & 66 & 66 \\
 & $\Omega2_{tau}$ & 0.5 & 0.5 & 0.5 \\ \hline
\multirow{3}{ 3em }{ $\Omega3$ } & $\Omega3_{mag}$ & 1 & 1 & 0.5 \\ 
 & $\Omega3_{offset}$ & 90 & 90 & 90 \\
 & $\Omega3_{tau}$ & 0.5 & 0.5 & 0.5 \\ \hline
  \multirow{3}{ 3em }{ $\Delta2$ } & $\Delta2_{mag}$ & 18 & 18 & 18 \\ 
 & $\Delta2_{offset}$ & 81 & 81 & 81 \\
 & $\Delta2_{tau}$ & 6 & 6 & 6 \\ \hline
 \multirow{3}{ 3em }{ $\Delta3$ } & $\Delta3_{mag}$ & 10 & 10 & 10 \\ 
 & $\Delta3_{offset}$ & 0 & 0 & 0 \\
 & $\Delta3_{tau}$ & 0 & 0 & 0 \\ \hline
 \multicolumn{2} { c | } { $ \gamma_f $} & 2 & 2 & 2 \\ 
 \multicolumn{2} { c | } { $ \gamma_{S1}$} & 20 & 20 & 20 \\
 \multicolumn{2} {c | } { $\gamma_{S2}$} & 20 & 20 & 20 \\
 \multicolumn{2}{c |}{$\gamma_{out}$} & 2 & 0.2 & 20 \\ \hline

\end{tabular}

\section{\label{sec:appB}Additional Graphs}

\begin{figure}[htbp]
\begin{center}
\includegraphics[width=0.95\linewidth]{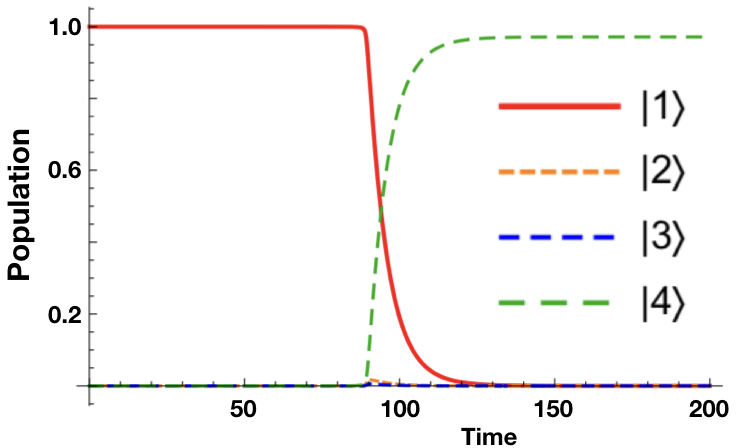}
\caption{Populations in the levels with $\gamma_{f}$ and $\gamma_{out}$ an order of magnitude smaller than the figures presented in the main body of the paper}
\label{fig:Decay_Weak}
\end{center}
\end{figure}

\begin{figure}[htbp]
\begin{center}
\includegraphics[width=0.95\linewidth]{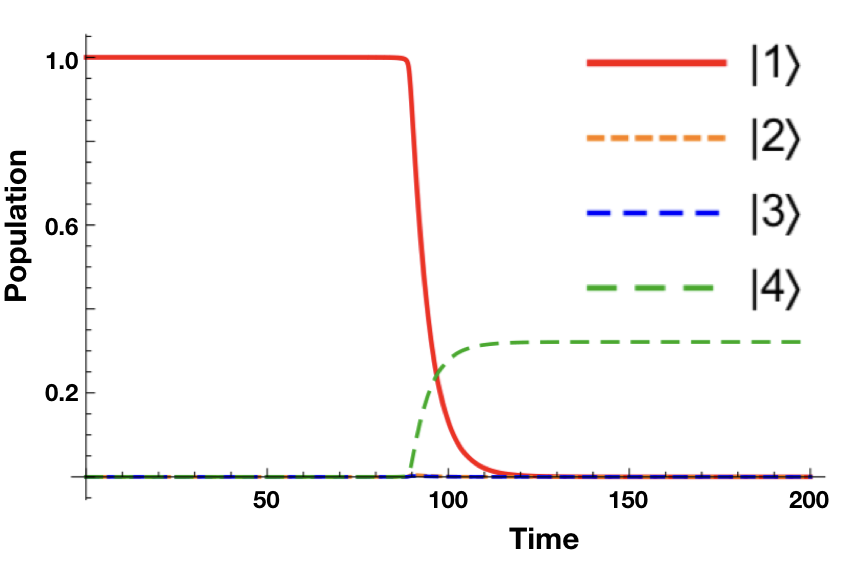}
\caption{Populations in the levels with $\gamma_{f}$ and $\gamma_{out}$ an order of magnitude larger than the figures presented in the main body of the paper}
\label{fig:Decay_Strong}
\end{center}
\end{figure}

\begin{figure}[htbp]
\begin{center}
\includegraphics[width=0.95\linewidth]{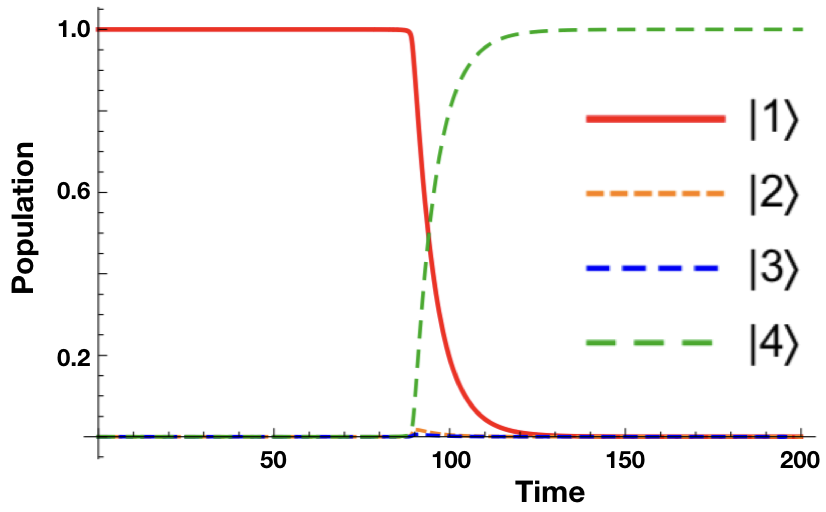}
\caption{This is a similar run to what is presented in \fig{fig:pops} but with $\gamma_{out}$ set to 0.}
\label{fig:InsideDecay}
\end{center}
\end{figure}

One thing of note from \fig{fig:InsideDecay} is how robust this procedure is with regards to the decay from $\vert 2 \rangle$ into $\vert 1 \rangle$.  The only thing that changes from \fig{fig:pops} is removing the decay from $\vert 2 \rangle$ to outside of the system, yet the improvement in performance is quite noticeable. 

\section{RAP and Feshbach Resonance energy structure}

\begin{figure}[htbp]
\begin{center} 
\includegraphics[width=0.95\linewidth]{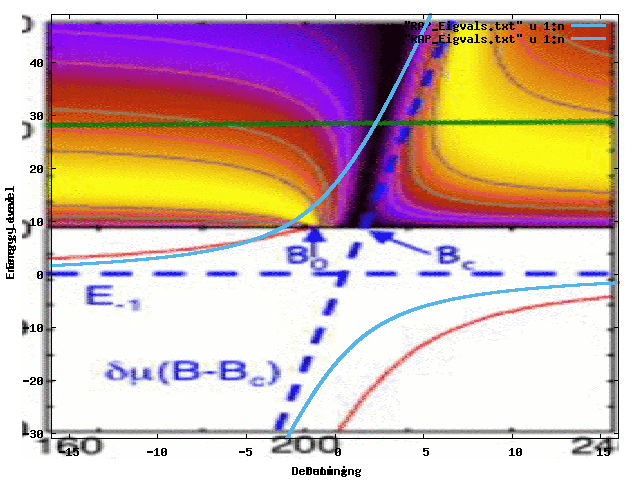}
\caption{Comparison of energies from Feshbach Resonance and Rapid Adiabatic Passage processes.  Our plot (blue lines) has time on the x-axis with Energy on the y-axis.  The background FR plot has B-field strength on the x-axis with energy on the y-axis.  From this we relate RAP level detuning to B-field detuning in FR.  Background image from \cite{FeshRes}.}
\label{fig:comp}
\end{center}
\end{figure}

\bibliography{MyBibliography}{}
\end{document}